\documentclass[prb,twocolumn,preprintnumbers,amsmath,amssymb,superscriptaddress]{revtex4}

\usepackage{graphicx}
\usepackage{dcolumn}
\usepackage{bm}
\usepackage{verbatim}

\begin{document}

\title{ How to Enhance Dephasing Time in Superconducting Qubits}

\author{{\L}ukasz Cywi{\'n}ski}
\affiliation{Condensed Matter Theory Center, Department of Physics, University of Maryland, College Park, MD 20742-4111, USA}
\author{Roman M. Lutchyn}
\affiliation{Condensed Matter Theory Center, Department of Physics, University of Maryland, College Park, MD 20742-4111, USA}
\affiliation{Joint Quantum Institute, Department of Physics, University of Maryland, College Park, MD 20742-4111, USA}
\author{Cody P. Nave}
\affiliation{Condensed Matter Theory Center, Department of Physics, University of Maryland, College Park, MD 20742-4111, USA}
\author{S. Das Sarma}
\affiliation{Condensed Matter Theory Center, Department of Physics, University of Maryland, College Park, MD 20742-4111, USA}
\affiliation{Joint Quantum Institute, Department of Physics, University of Maryland, College Park, MD 20742-4111, USA}
\date{\today}

\begin{abstract}
We theoretically investigate the influence of designed pulse sequences in restoring quantum coherence lost due to background noise in superconducting qubits. We consider both $1/f$ noise and Random Telegraph Noise, and show that the qubit coherence time can be substantially enhanced by carefully engineered pulse sequences. 
Conversely, the time dependence of qubit coherence under external pulse sequences could be used as a spectroscopic tool for extracting the noise mechanisms in superconducting qubits, i.e.~by using Uhrig's pulse sequence one can obtain information about moments of the spectral density of noise. 
We also study the effect of pulse sequences on the evolution of the qubit affected by a strongly coupled fluctuator, and show that the non-Gaussian features in decoherence are suppressed by the application of pulses.
\end{abstract}

\maketitle

\section{Introduction.}
Quantum decoherence, the continuous decay (``the loss of quantum memory'') of a quantum state due to its interaction with the environment provides the conceptual connection between the microscopic quantum and macroscopic classical worlds.\cite{Zurek_RMP03} Understanding and preventing decoherence is also central to the fledging field of quantum computation, as the loss of quantum coherence leads to errors in the processing of quantum information.
In fact, quantum error correction protocols, necessary for quantum computation, require the decoherence to be below a certain threshold.\cite{Nielsen_Chuang}
It is therefore of paramount importance that the decoherence of qubits, two-level systems used to store and process quantum information, is suppressed as much as possible. In this article, we develop realistic strategies, based on the application of designed external pulse sequences, that suppress an important source of decoherence in an important class of qubits, solid state superconducting qubits.\cite{Makhlin_RMP01, Devoret_Martinis_04} The decoherence mechanism considered in this work is that due to {\it classical noise}, {\it i.e.} a situation where the qubit couples to a random classical temporally fluctuating field. Such noise is, in fact, the major source of quantum dephasing in superconducting qubits,\cite{Nakamura_PRL02,Paladino_PRL02,Galperin_03,Makhlin_PRL04,Galperin_PRL06,Bergli_PRB06,Ithier_PRB05} and therefore the pulse sequences proposed in this work should be useful in restoring coherence in solid state superconducting quantum computer architectures.

For many decades, in the field of magnetic resonance, pulse sequence techniques have been studied as a method of reducing spin ensemble dephasing.\cite{Haeberlen,Vandersypen_RMP04} The most famous sequences are Hahn's spin echo, Carr-Purcell-Meiboom-Gill (CPMG) sequence,\cite{Haeberlen}  and periodic dynamical decoupling (PDD). 
Spin echo (SE) is the simplest, consisting of just one $\pi$ pulse, whereas CPMG is its multi-pulse generalization.The PDD sequence, introduced in the context  of quantum computation, was designed to average out the influence of the environment, effectively decoupling the qubit.\cite{Viola_PRA98,Viola_JMO04,Chen_PRA07} Dynamical decoupling was developed further by introducing an idea of  recursively defined sequences,\cite{Khodjasteh} termed concatenated dynamical decoupling (CDD). Recently, using a spin-boson model Uhrig obtained a new sequence\cite{Uhrig_PRL07} (termed here UDD), which nearly completely suppresses short time decoherence under certain conditions. UDD was later shown to be universal,\cite{Lee_UDD} in the sense that for {\it for any pure dephasing Hamiltonian} the $n$-pulse UDD sequence leads to the cancellation of $n$ orders of the time expansion of the off-diagonal element of the qubit density matrix $\rho_{_{+-}}(t)$.

In this paper, we study the effect of pulse sequences on the decoherence in superconducting (SC) qubits, subject to classical $1/f$ and Random Telegraph Noise (RTN). 
Experimental studies\cite{Nakamura_PRL02, Yoshihara_PRL06,Kakuyanagi_PRL07,Bialczak_PRL07} have shown that SC qubits suffer decoherence from $1/f$ noise, which is  associated with fluctuations of electric or magnetic dipoles in the insulating materials. 
In charge qubits, where the area of the tunnel junctions is small, it has been established that the qubit is often coupled to a few two-level fluctuators (TLFs), which can be treated as classical sources of RTN.\cite{Nakamura_PRL02, Ithier_PRB05, Paladino_PRL02, Galperin_PRL06}  Here we focus on the case of charge qubits, in which the charge noise is dominant, but our results, with minor modifications, are applicable to phase and flux qubits.

It has been experimentally shown that the coherence of the Cooper-pair box charge qubit is significantly prolonged by the application of the SE sequence.\cite{Nakamura_PRL02,Collin_PRL04,Ithier_PRB05} Characteristic plateaus seen in the echo signal have been explained theoretically as arising when the noise is dominated by a single {\it classical} TLF coupled to the qubit.\cite{Galperin_PRL06} 
Beyond SE, only PDD\cite{Martinis_PRB03,Shiokawa_PRA04,Faoro_PRL04,Falci_PRA04,Gutmann_PRA05,Bergli_PRB07} and CPMG\cite{Faoro_PRL04} sequences  have received theoretical attention in the context of SC qubits. In this work, we suggest the use of more sophisticated pulse sequences, such as CDD and UDD, to suppress noise-induced decoherence in SC qubits, finding that, depending on the details of the noise, CPMG or UDD is optimally effective in reducing decoherence in superconducting circuits. We emphasize that earlier work in the literature on CPMG,\cite{Witzel_PRL07} CDD,\cite{Yao_PRL07,Witzel_CDD_PRB07,Zhang_Viola_PRB08} and most recently UDD\cite{Lee_UDD} pulse sequences was carried out entirely in the context of electron spin decoherence in a nuclear spin bath, in which the quantum correlations within the bath are important. 
On the other hand, the bath fluctuations due to charge noise in superconducting qubits can often be treated classically as established in Ref.~\onlinecite{Galperin_PRL06}

Apart from prolonging the coherence time, pulse sequences could be used to gain valuable information about the environmental noise, as  the time dependence of decoherence is different for various sequences. 
The microscopic origin of the noise affecting the coherence and energy relaxation in the SC qubits is still a subject of ongoing research,\cite{Shnirman_PRL05,deSousa_PRL05,Faoro_PRL05,Faoro_Ioffe_PRL06,Constantin_PRL07,Lutchyn_08} and we discuss here  how pulse sequences can be used to learn more about the noise spectrum at low frequencies. This approach is complementary to using the measurement of energy relaxation time of the qubit for noise spectroscopy at higher frequencies, of the order of qubit energy splitting.\cite{Schoelkopf_spectrometer,Astafiev}

We consider here the experimentally relevant\cite{Nakamura_PRL02,Yoshihara_PRL06,Kakuyanagi_PRL07} situation in which decoherence is dominated by pure dephasing ({\it i.e.} ``$T_{2}$'') processes, and not by energy relaxation ({\it i.e.} ``$T_{1}$'') processes, $T_{1} \! \gg \! T_{2}$. 
In the current experiments pure dephasing is dominant mechanism of decoherence away from the so-called optimal bias point.\cite{Vion_Science02,Ithier_PRB05} The decoherence at the optimal point in present charge and flux qubit designs is limited by $T_{1}$ processes ({\it i.e.} $T_{2} \! \simeq \! 2T_{1}$), since the effect of noise is then suppressed to the first order.\cite{Makhlin_PRL04}  
However, the requirement of keeping the qubit at the optimal points at all times might be overly confining for a system of multiple interacting qubits.\cite{footnote_optimal} Thus, the ability to prolong the coherence of the qubit in the pure dephasing regime is still desirable.
Furthermore, in phase qubits\cite{Martinis_PRL02,Martinis_PRB03,Johnson_PRB03} there is no optimal point and the effect of pulse sequences should lead to a substantial increase of $T_2$ in the case when qubit coherence is $T_2$ limited.
Another strategy\cite{Koch_PRA07,You_PRB07} for suppressing the influence of the noise on the qubit was implemented recently.\cite{Schreier_07} In such a ``transmon'' qubit the coupling to the charge noise is exponentially suppressed. However, this qubit  is  still sensitive to the flux noise, the relevance of which becomes prominent away from the optimal flux bias point. 

The article is organized in the following way. In Sec.~\ref{sec:model} we introduce the pure dephasing Hamiltonian and describe the types of noise which we shall consider. Sec.~\ref{sec:sequences} contains the overview of various pulse sequences applicable to the pure dephasing case. In Sec.~\ref{sec:gaussian} we present the analytical solution for decoherence under pulses for the case of Gaussian noise, and we discuss how the pulse sequences act as filters suppressing the influence of low-frequency noise on the qubit dynamics. The calculations for Gaussian $1/f^{\alpha}$ noise are presented in Section \ref{sec:1f_gaussian}, where the role of the ultra-violet cutoff in the noise spectrum is highlighted. We also introduce the idea of using the UDD sequence to obtain the quantitative information about the low-frequency noise spectral density.
Finally, in Sec.~\ref{sec:RTN} we present the results for decoherence due to the RTN. We identify the regime in which the application of even a few pulses leads to the increase in the coherence time, and we find that the analytical Gaussian approximation to calculation of decoherence is asymptotically exact for large number of applied pulses.

\section{The Hamiltonian and the model of the noise} \label{sec:model}
The limit of energy relaxation time being much longer than the dephasing time, $T_{1} \! \gg \! T_{2}$, corresponds to using the pure dephasing Hamiltonian to describe the qubit-environment interaction: 
\begin{equation}
\hat{H}  =   \frac{1}{2}[\Omega + \beta(t)]\hat{\sigma}_{z} \,\,,   \label{eq:H}
\end{equation}
where $\Omega$ and $\beta(t)$ are, respectively, the qubit energy splitting and a classical random variable representing fluctuation of the energy splitting due to coupling to one or many TLFs. The function $\beta(t)$ represents a classical stochastic process, given by 
\begin{equation}
\beta(t)  =  \sum_{i}v_{i}\xi_{i}(t) \,\, ,
\end{equation}
where $\xi_{i}(t)\!\! = \!\! \pm 1/2$ corresponds to the RTN signal\cite{Machlup_JAP54} from  the $i$-th TLF, with $v_{i}$ being the corresponding coupling strength. 

The stochastic processes are defined by their correlation functions. The two-point correlation function is given by
\begin{equation}
S(t_{1}-t_{2}) = \langle \beta(t_{1})\beta(t_{2}) \rangle \,\, ,
\end{equation}
where $\langle ... \rangle$ is the average with respect to the noise realizations, and we have assumed here $\langle \beta(t) \rangle$$=$$0$. The Fourier transform of the two-point correlation function is the spectral density of noise (more generally referred to as the {\it first} spectral density, see Ref.~\onlinecite{Kogan}): 
\begin{equation}
S(\omega) = \int_{-\infty}^{\infty} e^{i\omega t} S(t)  dt \,\, .
\end{equation}
When the statistics of fluctuations are Gaussian, the noise is completely defined by $S(t)$, and the average over the noise realizations can be written as a Gaussian functional integral over all possible realizations of $\beta(t)$:
\begin{equation}
\langle ... \rangle = \int \mathcal{D}\beta \,\, e^{-\frac{1}{2} \int dt_{1} \int dt_{2} \beta(t_{1}) S^{-1}(t_{1}-t_{2})\beta(t_{2}) } ... \,\, ,  \label{eq:average}
\end{equation}
with $S^{-1}$ is defined by 
\begin{equation}
\int dt'' S^{-1}(t-t'')S(t''-t') = \delta(t-t') \,\, .
\end{equation}
On the other hand, when the noise is non-Gaussian, one has to consider also higher order correlation functions. For the relevant here case of the RTN, the two-point correlation function and its Fourier transform are given by
\begin{eqnarray}
S(t_{1}-t_{2}) & = & v^{2} \langle \xi(t_{1})\xi(t_{2}) \rangle = \frac{v^2}{4} e^{-2\gamma | t_{1}-t_{2}| } \,\, ,  \label{eq:S_RTN} \\
S(\omega) & = & \frac{\gamma v^2}{\omega^2 + 4\gamma^2} \,\, .  \label{eq:Lorentzian}
\end{eqnarray} 
where $\gamma$ is the rate of switching between the two values of $\xi\!\!=\!\!\pm 1/2$. 
We have used here the high temperature limit ($k_{\text{B}}T \! \gg \! \omega$) for the spectral function, {\it i.e.} the noise is symmetric,\cite{Machlup_JAP54,Galperin_PRL06} with both rates of transitions between the two states of the TLF being equal to $\gamma$.
If the noise had Gaussian statistics, one would have been able to express the higher order correlators through the two-point correlation function $S(t_{1}-t_{2})$.
This is not true for the RTN, and we refer interested reader to Appendix \ref{app:RTN} for more details.

When many TLFs with a log-uniform distribution of $\gamma$  ({\it i.e.} with probability of finding a fluctuator with a given $\gamma$ being $P(\gamma)\!\! \propto 1/\gamma$) contribute to $\beta(t)$, the spectral density is $S(\omega)\!\! \propto 1/\omega$. This is the well-known $1/f$ noise.\cite{Kogan} It extends to an infrared cutoff frequency $\omega_{\text{ir}}$ below which the spectrum flattens out, with values of $\omega_{\text{ir}}/2\pi \! < \! 1$ Hz in SC qubits.\cite{Ithier_PRB05} The log-normal distribution of switching rates arises when $\gamma$ depends exponentially on another quantity having a uniform distribution. For example, in a model of localized TLFs one obtains a log-uniform distribution of tunnel splittings which depend exponentially on the tunnel barrier height,\cite{Shnirman_PRL05} and in the recently proposed model of the Andreev fluctuator bath\cite{Faoro_PRL05,Lutchyn_08} the switching rate of the effective TLF depends exponentially on the distance between the pair of impurity sites participating in the Andreev tunneling process.

\section{Decoherence under pulse sequences.}  \label{sec:sequences}
We consider the decoherence of the qubit prepared initially in the coherent superposition of its ``up'' and ``down'' states. Specifically, at time $t'\!\!=\!\! 0$ we assume that the qubit's state vector is $|\Psi(0)\rangle \!\! = \!\! a|\uparrow \rangle + b|\downarrow \rangle$ with $|a|^2\!\! = \!\! |b|^{2} \!\! = \!\! 1/2$, which in the Bloch vector language corresponds to the vector being in the $xy$ plane. Experimentally this is achieved by initializing the qubit in one of the eigenstates of $\hat{\sigma}_{z}$ and applying a $\pi/2$ rotation about $x$ or $y$  axis at initial time.\cite{Makhlin_RMP01}

In Free Induction Decay (FID) experiment we let the qubit evolve freely for time $t$, and then perform a measurement. Due to the noise in the Hamiltonian (\ref{eq:H}), the qubit state at the measurement time is
\begin{equation}
|\Psi(t)\rangle = e^{-i\Omega t/2}e^{-\frac{i}{2}\int_{0}^{t}\beta(t')dt'} a |\uparrow\rangle + e^{i\Omega t/2}e^{\frac{i}{2}\int_{0}^{t}\beta(t')dt'}  b|\downarrow\rangle \,\, , 
\end{equation}
so that the off-diagonal element of the qubit density matrix is (we use the units with $\hbar \! = \! 1$)
\begin{equation}
\rho^{\text{FID}}_{+-}(t) =  e^{-i\Omega t}e^{-i\int_{0}^{t}\beta (t')dt'}  \rho_{+-}(0) \,\, 
\end{equation}
where $\rho_{+-}(0)\!\! = \!\! ab^{*}$.
We quantify the qubit coherence using function $W(t)$, defined in the following way:
\begin{equation}
W(t) \equiv \frac{| \langle \rho_{+-}(t)\rangle |}{|\langle \rho_{+-}(0) \rangle |} \,\, .
\end{equation}

We are now going to consider applying a certain number $n$ of {\it ideal} ($\delta$-shaped) $\pi$ pulses (about, for example, $x$ axis) in the time interval $t'\in [0,t]$. In the following, $t$ will always denote the measurement time, and by $W(t)$ we mean the coherence at time $t$ with $n$ pulses applied {\it within this time}. 

\begin{figure}
\includegraphics[width=8.5cm]{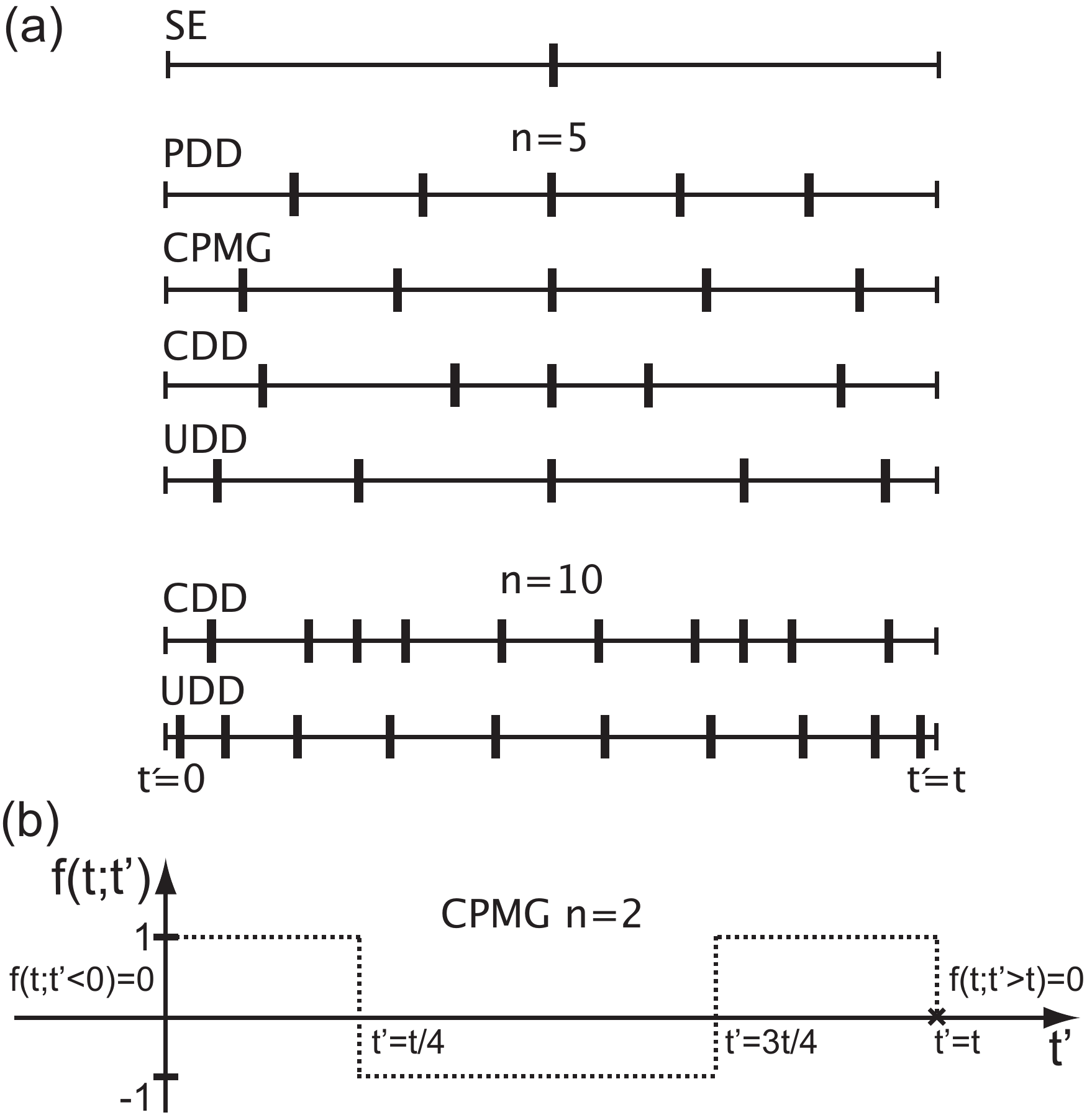}
\caption{(a) The illustration of various pulse sequences with application times of $\pi$ pulses marked. Spin echo (SE) is shown along with the PDD, CPMG, CDD and UDD sequences with $n\!\!=\!\! 5$ pulses (for CDD this corresponds to the $3^{rd}$ order of concatenation). 10 pulse UDD and CDD ($4^{th}$ order of concatenation) are also shown. (b) Function $f(t;t')$ defined in Eq.~(\ref{eq:f}) for 2 pulse CPMG sequence. } \label{fig:sequences}
\end{figure}

Since the $\pi$ rotation about the $x$ axis is given by $\exp(-i\pi \hat{\sigma}_{x}/2) \! = \! -i\hat{\sigma}_{x}$,  the qubit evolution operator with $\pi$ pulses applied at times $t_{1},...,t_{n}$ is
\begin{align}
\hat{U}(t) &= e^{-\frac{i}{2}\hat{\sigma}_{z} \int_{t_{n}}^{t} [\Omega+ \beta(t')]dt'} (-i\hat{\sigma}_{x}) e^{-\frac{i}{2}\hat{\sigma}_{z} \int_{t_{n-1}}^{t_{n}} [\Omega+\beta(t')]dt'} ...  \nonumber \\
& ...(-i\hat{\sigma}_{x}) e^{-\frac{i}{2}\hat{\sigma}_{z} \int_{0}^{t_{1}} [\Omega+\beta(t')]dt'}  \,\, .
\end{align}
The $\hat{\sigma}_{x}$ operators exchange the amplitudes of $|\uparrow\rangle$ and $|\downarrow\rangle$ states of the qubit, and we arrive at the decoherence function under the action of the pulse sequence:
\begin{equation}
W(t) = \left | \left\langle \exp ( -i\int_{0}^{t} \beta(t') f(t;t') dt' ) \right\rangle \right | \,\, .  \label{eq:Wpulses}
\end{equation}
In this equation we have introduced the function $f(t;t')$ which characterizes the pulse sequence:
\begin{equation}
f(t;t') = \sum_{k=0}^{n} (-1)^{k} \Theta(t_{k+1} - t')\Theta(t'-t_{k}) \,\, ,  \label{eq:f}
\end{equation}
where $\Theta(t')$ is the Heaviside step function, $t_{0}\! = \! 0$ and $t_{n+1}\! = \! t$, the total evolution time. This function switches between $1$ and $-1$ at the times at which the $\pi$ pulses are applied and $f(t;t')\! = \! 0$ for $t'<0$ and $t'>t$. In Fig.~\ref{fig:sequences}b we show, as an example, a plot of $f(t;t')$ for 2 pulse CPMG sequence.

We denote the characteristic time of decay of $W(t)$ as $T_{2}$, defined by $\ln W(T_{2}) \!\! = \!\! -1$. It depends on the pulse sequence applied during the qubit evolution, and this dependence on the number of pulses $n$ and their spacing is the main subject of this paper. In most of the cases considered here the decoherence is $\it not$ described by a simple exponential decay $W(t)\!\! \sim \exp(-t/T_{2})$. Such a decay law appears when the relevant dynamical  time-scale of the environment ({\it i.e.} the noise autocorrelation time) is much smaller than $T_{2}$ (the Markovian limit of qubit dynamics. This is not true for $1/f$ noise which is correlated on a very long time-scale, and also for the RTN due to a slow fluctuator (with small $\gamma$). 

\subsection{Pulse sequences aimed at suppressing the pure dephasing}
We denote the times at which the $n$ pulses are applied by $t_{k}\!\!=\!\!\delta_{k}t$ with $0\!\!\leq\!\!\delta_{k}\!\!\leq\!\!1$, with $k\! = \! 1 ... n$. The spacing of times $t_{k}$ in the sequences under consideration here is illustrated in Fig.~\ref{fig:sequences}a.
Application of a single $\pi$ pulse at $t'\!\!= \!\! t/2$ ($n\!=\! 1$, $\delta_{1}\! = \! 1/2$) corresponds to  the  spin echo (SE) experiment. The $\pi$ pulse exchanges the amplitudes of the two states of the qubit, and the evolution during the remaining time period partially cancels the evolution before the pulse. More specifically, the echo sequence removes the influence of the noise frequencies $\omega$ smaller than $1/t$ (the quasi-static shifts of the qubit precession frequency). This, of course, gives the complete cancellation of the {\it static} randomness in qubit splittings $\Omega$ in the measurement of an ensemble of qubits (inhomogeneous broadening). In the considered here case of the SC qubits, one always deals with repeated measurements of a single qubit (a time ensemble), and SE is a very efficient technique which suppresses the low-frequency part of the $1/f$ noise, leading to a substantial increase in the $T_{2}$ time 
in superconducting charge\cite{Nakamura_PRL02,Collin_PRL04,Ithier_PRB05} and flux qubits.\cite{Yoshihara_PRL06,Kakuyanagi_PRL07}

The multiple-pulse extension of the echo  is the Carr-Purcell-Meiboom-Gill (CPMG) sequence,\cite{Haeberlen} defined by $\delta_{k}\!\! = \!\!(k  - 1/2)/n$. This sequence leads to periodic echo-like revivals of the coherence. While it was used for many years in NMR experiments performed on ensembles of spins, it was recently predicted that it should be highly effective at prolonging the coherence of a spin qubit interacting with the nuclear bath.\cite{Witzel_PRL07}
Whereas CPMG is best understood as a {\it refocusing} sequence, in recent years a lot of attention has been devoted to the idea of {\it dynamical decoupling} of the qubits from its environment by pulses.\cite{Viola_PRA98,Viola_JMO04,Khodjasteh,Chen_PRA07} In this approach the pulses are used to average out the influence of the environment on the qubit, which can be fully achieved only in the limit of very fast repetition of pulses. Out of many types of dynamical decoupling (DD) protocols, we concentrate here on deterministic periodic (PDD) sequence (see {\it e.g.}~Ref.~\onlinecite{Zhang_Viola_PRB08} for a comparison of more kinds of DD techniques applied to the spin bath problem). It is defined by $\delta_{k}$$=$$k/(n+1)$. Although it looks very similar to CPMG (the difference being only a small offset of the initial and final delay times), below we will show that CPMG visibly outperforms PDD when considering the realistic small $n$. The key difference is that while in the limit of very fast application of pulses both PDD and CPMG {\it decouple} the qubit from the environment (during the whole time of the evolution), for realistic small $n$ the CPMG sequence is much better at {\it refocusing} the coherence at the final time $t'\! = \! t$. 

Recently, a new family of DD protocols involving concatenating (recursively embedding the sequences within themselves) has been proposed.\cite{Khodjasteh} For the purpose of combating the pure dephasing, we will concentrate on concatenations of the echo sequence, and for simplicity we will refer to it as simply CDD. The CDD sequence at $l$-th order of concatenation and for total evolution time $t$ is defined as $\mathrm{CDD}_{l}(t)$. $\mathrm{CDD}_{0}(t)$ is free evolution for time $t$.
The $l$-th order of concatenation is then recursively defined by 
\begin{equation}
\mathrm{CDD}_{l}(t) \equiv \mathrm{CDD}_{l-1}\left (\frac{t}{2} \right) - \pi - \mathrm{CDD}_{l-1}\left (\frac{t}{2} \right )  - \pi \,\, ,
\end{equation}
so that, $\mathrm{CDD}_{1}(t)$ is the SE with a $\pi$ pulse at $t/2$ and $\mathrm{CDD}_{2}(t)$ is the same as $n\! = \! 2$ CPMG sequence. For $l \! > \! 2$ the concatenations of the echo give us new sequences of non-trivially spaced $\pi$ pulses, the performance of which has been investigated theoretically in the case of the nuclear spin bath.\cite{Yao_PRL07,Witzel_CDD_PRB07,Zhang_Viola_PRB08} Note that the number of pulses for the $l$-th order of concatenation is $n \! \approx \! 2^{l}$.
The CDD sequences were argued\cite{Khodjasteh} to be more tolerant to implementation errors and more efficient (in terms of performance for the same number of pulses) than the PDD sequence. However, the theoretical comparisons\cite{Khodjasteh} with other DD protocols were done in the quantum mechanical setting, using the Magnus expansion of the evolution operators, or so-called ``average Hamiltonian'' theory.\cite{Vandersypen_RMP04} CDD is designed to cancel, with each order of concatenation, successive orders of the qubit-bath interaction in the Magnus expansion. It also cancels successive orders of intra-bath interaction.\cite{Witzel_CDD_PRB07}
However, it is not {\it a priori} clear whether the advantages of CDD are going to also hold for the case of dephasing due to classical noise.

The most recent development in suppressing the pure dephasing was the introduction of a new sequence by Uhrig,\cite{Uhrig_PRL07,Uhrig_08} which we term here UDD. This sequence was optimized for pure dephasing due to a bosonic environment or classical Gaussian noise, but later its surprising universal character was discovered in a general quantum-mechanical setting.\cite{Lee_UDD} 
UDD is defined by
\begin{equation}
\delta_{k} =  \sin^{2}[\pi k/(2n+2)] \,\, , \label{eq:delta_UDD}
\end{equation}
and in the next Section we will explain in what sense it is ``optimal'' for the case of the Gaussian noise.
Originally the sequence was applied\cite{Uhrig_PRL07} to the case of the environment characterized by an Ohmic spectral density of the noise having a sharp high-frequency cutoff, $S(\omega) \sim \omega \Theta(\omega_{c}-\omega)$, with the ultra-violet cutoff $\omega_{c}$. Here we will analyze its performance for Gaussian noise with $1/f$ spectral density (with and without the ultra-violet cutoff in the spectrum), and for classical {\it non-Gaussian} RTN.

\subsection{Realistic pulses}  \label{sec:realistic}
We consider here the case of ideal, {\it i.e.} $\delta$-shaped $\pi$ pulses (so-called ``bang-bang'' or unbounded control). In reality, the pulses will have a finite duration $\tau_{p}$ and they might be imperfect, {\it e.g.} one can have pulse length or amplitude errors (leading to a wrong angle of rotation) or an off-resonance error, due to which the rotation occurs around a tilted axis. When these errors are systematic, they can be suppressed by using composite pulses.\cite{Cummins_PRA03} Pulse shaping has also been used to counteract the effect of the bath noise during the finite $\tau_{p}$ of the realistic pulse. Shapes of finite duration  $\pi$ and $\pi/2$ pulses were optimized to cancel the lowest order (in $\tau_{p}$) corrections due to interaction with arbitrary bath.\cite{Pasini_PRA08,Pryadko_PRA08} Optimization of control pulses was also considered for qubit coupled to a source of classical RTN\cite{Mottonen_PRA06} or classical $1/f$ noise,\cite{Kuopanportti_PRA08} and qubit interacting with a quantum two-level system.\cite{Rebentrost_06} 

These works show that the realistic $\pi$ pulses can be made quite robust to both implementation errors and environmental noise, and treating them as $\delta$-shaped is a good approximation as long as $\tau_{p}$ is larger than $\tau_{\text{min}}$, the minimal interval between the pulses in a given sequence. For UDD this time scales with the number of pulses $n$ as  $\tau_{\text{min}} \! \propto t/n^2$ in contrast to $t/n$ scaling for all the other sequences under consideration, see Fig.~\ref{fig:sequences}a for example with $n\! = \! 10$. In reality there is a lower limit on $\tau_{p}$, related to the presence of higher energy levels in the full spectrum of the system. The inevitable higher-order pulse errors (unaccounted for by optimization) can also add up in a sequence with large $n$.
Let us also mention that while CDD was shown\cite{Khodjasteh} to be robust against certain types of pulse errors by construction (the systematic errors being cancelled by successive concatenations), such an investigation has not been made in the case of UDD. It is not known how sensitive is the performance of this sequence to, {\it e.g.}, errors in timing of the pulses, which have to be spaced in a quite intricate fashion. 
These considerations lead us to concentrate on the case of rather small $n$. Instead of looking at a ``stroboscopic'' limit of dynamical decoupling of the qubit by fast repetition of pulses, we start from the echo sequence and show how the decoherence changes as we increase $n$ from one to ten. 

\section{Gaussian noise.}   \label{sec:gaussian}
We write the  decoherence function $W(t)$ from Eq.~(\ref{eq:Wpulses}) as
\begin{equation}
W(t) \equiv e^{-\chi(t)} \,\, ,
\end{equation}
defining the function $\chi(t)$.
In the Gaussian approximation, the average over noise can be performed using Eq.~(\ref{eq:average}), and $\chi(t)$ can be expressed through the spectral density of the noise $S(\omega)$ as
\begin{equation}
\chi(t) =  \int_{0}^{\infty} \frac{d\omega}{2\pi} S(\omega) |\tilde{f}(t;\omega)|^2 = \int_{0}^{\infty} \frac{d\omega}{\pi} S(\omega) \frac{F(\omega t)}{\omega^{2}} \,\, ,  \label{eq:chi_gaussian}
\end{equation}
where $\tilde{f}(t;\omega)$ is the Fourier transform of $f(t;t')$
with respect to $t'$. The filter function $F(\omega t)\! = \! \frac{\omega^2}{2}  |\tilde{f}(t;\omega)|^2$ encapsulates
the influence of the pulse sequence on decoherence.\cite{Martinis_PRB03,Sousa_review} In terms of times $t_{k}$ at which the pulses are applied (with $t_{0}\! = \! 0$ and $t_{n+1}\! = \! t$) we have
\begin{equation}
F(\omega t) =  \frac{1}{2}\Big| \sum_{k=0}^{n} (-1)^{k} (e^{i\omega t_{k+1}} - e^{i\omega t_{k}} ) \Big|^{2} \,\, . \label{eq:Fsum}
\end{equation}
Analytical expressions for $F(\omega t)$ for the sequences under consideration are given in Table \ref{tab:F}. Let us note the existence of the following sum rule for the filter functions:
\begin{equation}
\int_{-\infty}^{+\infty} \frac{d\omega}{\pi} \frac{F(\omega t)}{\omega^{2}} = \int_{-\infty}^{+\infty} f^{2}(t;t')dt'=t \,\, . \label{eq:sumrule}
\end{equation}
From this one can see that the pulse sequences cannot prolong the coherence time when the integral in Eq.~(\ref{eq:chi_gaussian}) is dominated by an initially flat $S(\omega)\! \simeq \! S(0)$ at low frequencies. We obtain then $\chi(t) \! = \! S(0)t/2$ for all pulse sequences at times $t \gg 1/\omega_{f}$ with $\omega_{f}$ being the frequency at which the noise spectrum starts to decay.

\subsection{The filter functions}
For free-induction decay (FID), as $\omega\rightarrow0$, $F(\omega t)/\omega^{2} \rightarrow t^{2}/2$. As a result, low frequency noise significantly contributes to $\chi(t)$, and for $1/f$ noise with $S(\omega)\! = \! A_{0}^2/\omega$ we get\cite{Schriefl_NJP06}
\begin{equation}
\chi(t) \approx \frac{(A_{0}t)^2}{2\pi} \ln \frac{1}{\omega_{\text{ir}}t} \,\, , 
\end{equation}
where $\omega_{\text{ir}}$ it the infra-red cutoff of the $1/f$ noise. In current experiments the microscopic cutoff is not reached, and $\omega_{\text{ir}}$ is determined by the measurement procedure, {\it i.e.} $\omega_{\text{ir}} \! \sim t_{\text{m}}$, where $t_{\text{m}}$ is the averaging time. Values of $\omega_{\text{ir}}/2\pi \! \sim 1$ Hz have been reported for SC charge qubits.\cite{Ithier_PRB05}
This exposure to small-$\omega$ noise is already removed with the SE sequence, for which $F(\omega t) \! \sim \! (\omega t)^4$ for $t \ll 4/\omega$. For $1/f$ noise this leads to a significant (by at least an order of magnitude) increase of the observed $T_{2}$ time in comparison to the FID experiment.\cite{Nakamura_PRL02,Ithier_PRB05,Yoshihara_PRL06,Kakuyanagi_PRL07} 

The PDD filter for $\omega \! < \! 2/t$ is $F^{\text{PDD}}(\omega t) \! \sim \! \omega^{4} \, (\omega^2)$ for odd (even) $n$ (see Table \ref{tab:F}), so that only odd $n$ sequence can suppress low-frequency noise. For larger frequencies, but smaller than $2n/t$ we have $F^{\text{PDD}}(\omega t) \! \sim \! (\omega t)^{2}/(2n+2)^2$.
On the other hand, the CPMG filter is propotional to $\omega^4 \, (\omega^{6})$ for odd (even) $n$, suppressing low frequency noise with $S(\omega)\! \sim \! 1/\omega^{\alpha}$ with $\alpha \! < \! 2 \, (4)$. Furthermore, for $\omega \! < 2n/t$ we have  $F^{\text{CPMG}}_{n}(\omega t) \! \sim \! (\omega t)^{4}/(2n)^{4}$. A small change of the initial and final interval between the pulses in comparison to PDD, leads to a more efficient high-pass filter of the noise.

\begin{figure}
\includegraphics[width=8.5cm]{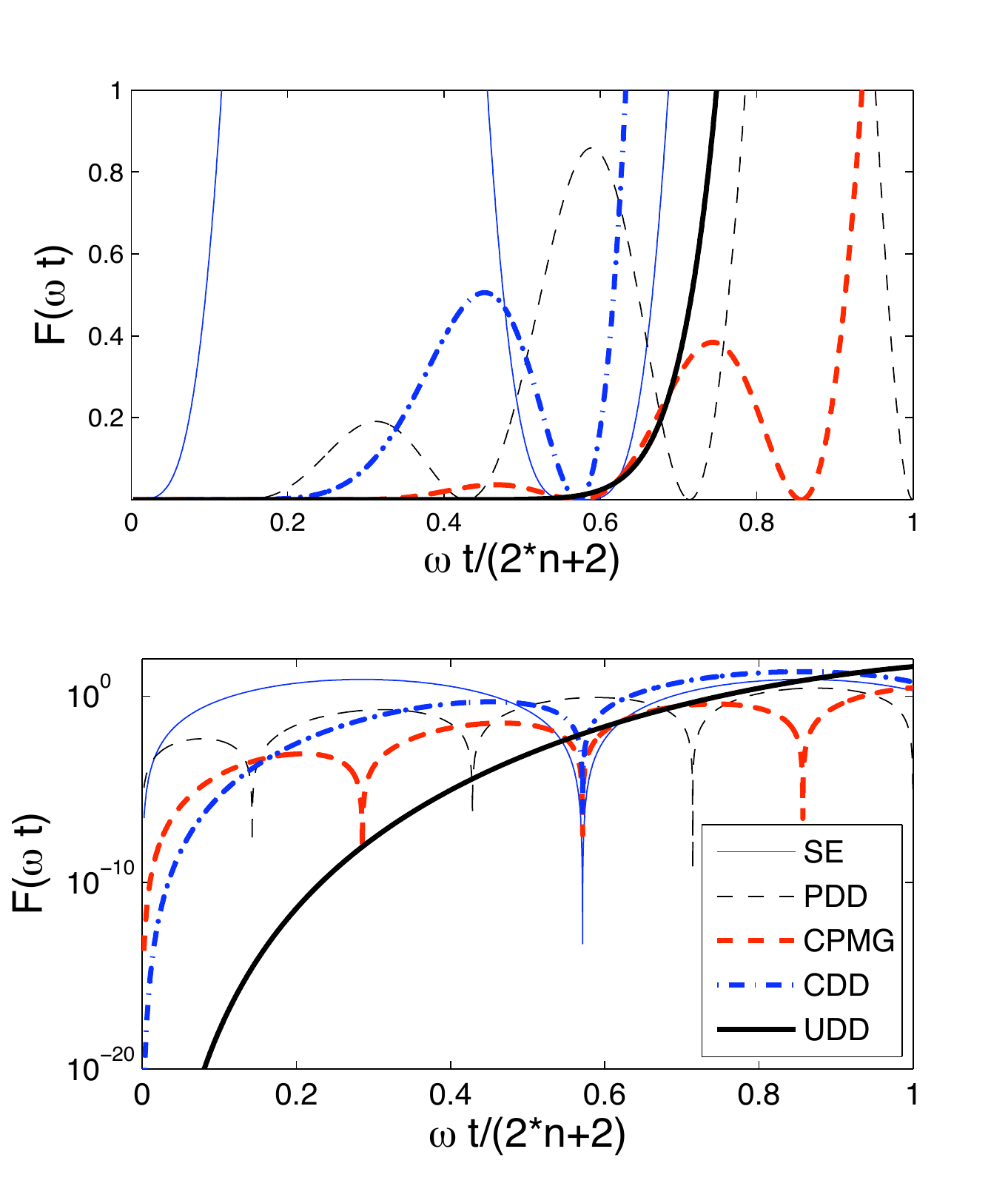}
\caption{(Color online) The filter functions $F(\omega t)$ for SE and various pulse sequences with $n\! = \! 10$ pulses (for CDD it corresponds to the 4th level of concatenation).The lower panel shows $F(\omega t)$ in the logarithmic scale. 
} \label{fig:F}
\end{figure}

From the recursive definition of CDD we get in the $l$-th order of concatenation
\begin{equation}
\tilde{f}^{(l)}(\omega) = \frac{1}{2}\tilde{f}^{(k-1)}\left(\frac{\omega}{2}\right )(1-e^{i\omega t/2} ) \,\, ,
\end{equation}
with $\tilde{f}^{(0)}$$=$$i/\omega(1-e^{i\omega t})$. From this the formula for the filter $F^{\text{CDD}}_{l}(\omega t)$ given in Table \ref{tab:F} follows. For frequencies $\omega \! < \! 4/t$ we have
\begin{equation}
F^{CDD}_{l} \approx \frac{ (\omega t)^{2l+2}}{2^{(l+1)^2+1}}  \,\, .
\end{equation}
It is important to note that unlike in the case of the other pulse sequences, the frequency at which $F(\omega t)$ becomes larger than $1$ scales not as $n/t$, but as $\sqrt{n}/t$ (for large $n$). This is illustrated in Fig.~\ref{fig:F}, where the filters $F(\omega t)$ are shown for $n\! = \! 10$ for all the sequences under consideration. The CDD filter is the first to become large with increasing $\omega$. The advantage of CDD over much simpler PDD and CPMG sequences is apparent only at very low frequencies - but it is the UDD filter which is clearly the best for $\omega \! \ll \! 2n$, see Fig.~\ref{fig:F}.

\begin{table}[t]
\begin{ruledtabular}
\begin{tabular}{cc}
Sequence &  $F(z)$   \\
\hline
FID                 & $ 2\sin^{2}\frac{z}{2}$ \\
SE              & $ 8\sin^{4}\frac{z}{4}$ \\
PDD  (odd $n$)       & $ 2\tan^{2}\frac{z}{2n+2} \sin^{2}\frac{z}{2}$ \\
CPMG (even $n$) & $ 8\sin^{4}\frac{z}{4n}\sin^{2}\frac{z}{2} / \cos^{2}\frac{z}{2n}$ \\
CDD             & $ 2^{2l+1} \sin^{2}\frac{z}{2^{l+1}} \prod_{k=1}^{l} \sin^{2}\frac{z}{2^{k+1}} $ \\
UDD             & $\frac{1}{2}\Big|  \sum_{k=-n-1}^{n} (-1)^{k} \exp[\frac{iz}{2}\cos \frac{\pi k}{n+1} ] \Big|^2$ \\
\end{tabular}
\end{ruledtabular}
\caption{The expressions for filter functions for various pulse
sequences. Here $n$ is the number of pulses, and $l$ is the order of
concatenation for CDD ($n$$\approx$$2^{l}$). In the range of $z\! <
\! 2n$, the filter function for UDD is very small, see Ref.~\onlinecite{Uhrig_PRL07}.
In the formulas for even-$n$  PDD and odd-$n$ CPMG, $\sin^{2}(z/2)$ is replaced with $\cos^{2}(z/2)$.
} \label{tab:F}
\end{table}

The $n$ pulse UDD sequence is derived by optimizing $\chi(t)$, choosing $\delta_{k}$ so that the first $2n+1$ terms in time expansion of $\chi(t)$ around $t\!\!=\!\! 0$ are zero (\emph{i.e.} setting the first $n$ terms in time expansion of $\tilde{f}(t;\omega)$ to zero). The physically meaningful solution\cite{Uhrig_PRL07,Uhrig_08} to the resulting set of nonlinear equations is given by Eq.~(\ref{eq:delta_UDD}). The resulting $F(\omega t)$ is  
\begin{equation}
F^{\text{UDD}}_{n}(\omega t) \approx  \frac{8(n+1)^2}{[(n+1)!]^2} \Big( \frac{\omega t}{4} \Big)^{2n+2} \,\, \text{for} \,\, \omega< 2/t \,\, , \label{eq:FUDD_loww}
\end{equation}
and $F^{\text{UDD}}_{n} \! \sim \! [\omega t/(2n+2)]^{2n+2} \ll 1$ for $\omega \ll (2n+2)/t$. 
Of all sequences considered here, the UDD gives the filter which most strongly suppresses the noise at low frequencies, as shown in Fig.~\ref{fig:F}.

Summarizing, the application of $n$ pulses within time $t$ effectively suppresses the noise power below frequency $\omega_{n}\!\! \sim \!\! 2n/t$, with the UDD sequence being, by construction, the most efficient high-pass filter. Thus, at short time $t$ or for large $n$ only the high-frequency fluctuations with $\omega \! > \! 2n/t$ contribute to $\chi(t)$. 

\subsection{Gaussian $1/f^{\alpha}$ noise}  \label{sec:1f_gaussian}
\begin{figure}
\includegraphics[width=9cm]{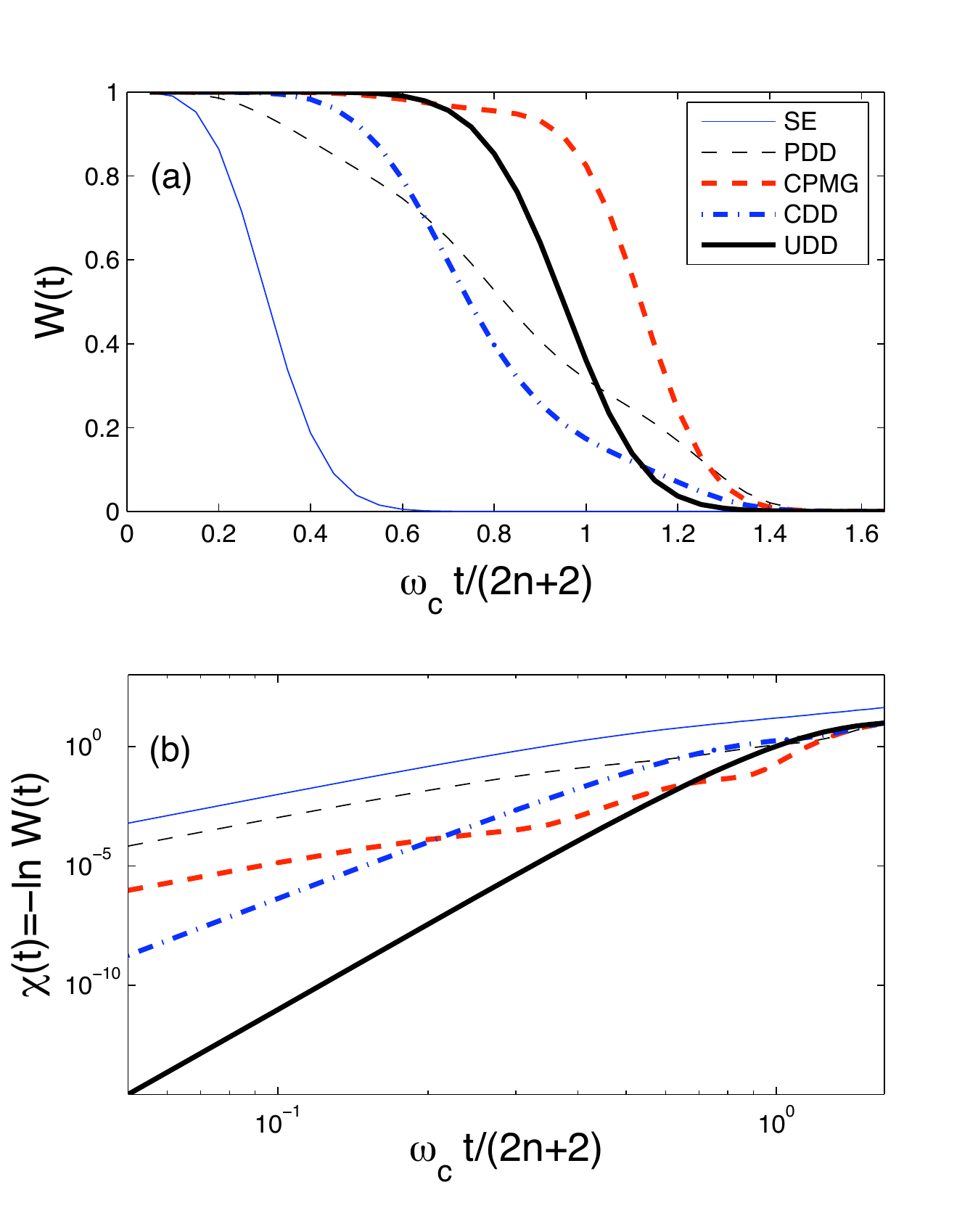}
\caption{(Color online) The dependence of (a) $W(t)$ and (b) $\chi(t) \!\! = \!\! - \ln W(t)$ for SE and higher-order ($n\!=\! 5$) sequences for $1/f$ noise with $A_{0}/\omega_{c}\! = \! 1$. UDD and CPMG give $W(t) \! \approx \! 1$ for $t\! < \! (2n+2)/\omega_{c}$. Echo signal (SE) (equivalent to PDD or CPMG with $n$$=$$1$) is also shown for comparison.
} \label{fig:cutoff}
\end{figure}

We concentrate now on the case of Gaussian $1/f^{\alpha}$ noise with spectral density $S(\omega)$$=$$A_0^{1+\alpha}/\omega^{\alpha}$, where $0.5 \leq \alpha \leq 1.5$. The conditions under which the $1/f$ noise originating from multiple TLFs is Gaussian are discussed in Ref.~\onlinecite{Schriefl_NJP06}.
We first consider the case in which an
ultra-violet cutoff $\omega_{c}$ is present in the noise spectrum, as it was inferred for charge noise from experiments in Ref.~\onlinecite{Ithier_PRB05}.
If we apply $n$ pulses in time $t$ such that $2n/t$$>$$\omega_{c}$, all the noise is strongly suppressed, as shown see Fig.~\ref{fig:cutoff}. Observation of an initially flat $W(t)$ is a clear-cut signature of a finite cutoff. Therefore, pulse sequences can provide important insight into the noise spectrum. 

The decay of qubit coherence for various pulse sequences is shown in Fig.~\ref{fig:cutoff}, where we compare $W(t)$ for various 5-pulse sequences with the echo (SE). The FID signal is not shown, since it depends on measurement-specific infra-red cutoff $\omega_{\text{ir}}$. However, in typical experimental situations $W^{\text{FID}}$ decays much faster than $W^{\text{SE}}$. 
For a given $n$, PDD is clearly the least effective approach at all times. As expected, for short times $t\!\! \ll \!\! 2 n / \omega_c$, UDD is orders of magnitude better than the other pulse sequences, see Fig.~\ref{fig:cutoff}b. Thus, UDD is the ideal sequence for maintaining a low level of decoherence, {\it i.e.} high fidelity, which is a necessary condition for quantum error correction. However, if the goal is simply to increase the characteristic decoherence time $T_{2}$, defined by $\chi(T_{2})\!\!=\!\!1$, then for a given $n$, the CPMG sequence is the best strategy. 

It is interesting to note that the CDD sequence does not offer strong advantages compared to CPMG and UDD. At short times it gives smaller $\chi(t)$ compared to CPMG, but the difference is not as dramatic as in the case of UDD. At longer times, CPMG is better and gives larger $T_{2}$. It seems that the benefits of using CDD expected in the regime of quantum bath dynamics are largely lost when dealing with classical noise.

A different situation arises when the cutoff cannot be reached, {\it i.e.}~we cannot apply pulses fast enough.
Then, only the low-frequency part of $S(\omega)$ can be suppressed, and the decay of $W(t)$ is due to high-frequency tail of noise spectrum.
For all the sequences under consideration it can be shown that to a very good approximation (although slightly worse for PDD) the integration of Eq.~(\ref{eq:chi_gaussian})  gives
\begin{equation}
\chi(t) \simeq \frac{C_{\alpha}}{2 \pi} \frac{(A_0 t)^{1+\alpha}}{n^{\alpha}} \,\, . \label{eq:chi1f}
\end{equation}
Here $C_{\alpha}$ is a sequence-specific constant of the order of one. This result can be quickly established  for CPMG in the following way. Its filter function $F(z\!=\!\omega t)$ consist of a periodic train of narrow peaks, with periods $z_{p} \! = \! 2\pi n$. 
We approximate these peaks by square steps of width $\Delta z$, which can be derived from the sum rule for $F(z)$ (Eq.~(\ref{eq:sumrule})). Then, to the lowest order in small quantity $\Delta z/z_{p} \! = \! 1/n$ we get for that for CPMG we have
\begin{equation}
C_{\alpha} = \frac{1}{\pi^{2}(2\pi)^{\alpha -1}} \sum_{k=1}^{\infty} \frac{1}{(k-1/2)^{\alpha}} \,\, ,
\end{equation}
which for $\alpha \! = \! 1$ gives $C_{1} \! = \! 7\zeta(3)/\pi^2 \! \simeq \! 0.85$ (where $\zeta(z)$ is the Riemann zeta function). 
On the other hand, for UDD one can use a seemingly crude approximation of $F^{\text{UDD}}_{n}(z) \! = \! a_{n} \Theta(z-(2n+2))$, with a constant $a_{n}$. This is motivated by the fact that the UDD filter for $z\! > \! 2n+2$ looks similar to a random signal, since it is given by a sum of periodic functions with non-commensurate periods (see Eqs.~(\ref{eq:delta_UDD}) and (\ref{eq:Fsum})). From Eq.~(\ref{eq:sumrule}) we get $a_{n} \!\ = \! \pi(n+1)$, and this value gives a good agreement with results of numerical integration using the exact form of the filter. The analytical approximation gives then $C_{\alpha} \! = \! \pi^2/[(\alpha+1)2^{\alpha +1}]$ and $n$ should be replaced by $n+1$ in the denominator in Eq.~(\ref{eq:chi1f}).
From these formulas, and from numerical calculations confirming their accuracy, we find that CPMG marginally outperforms the other sequences (more visibly for larger values of $\alpha$), so it is enough to implement this simple sequence to prolong qubit coherence in this regime.

\subsection{Noise spectroscopy using the pulse sequences}
The time dependence of qubit coherence under external pulse sequences can be used as a spectroscopic tool for extracting the noise spectrum contributing to dephasing. The idea of using qubit energy relaxation for noise spectroscopy was introduced in Ref.~\onlinecite{Schoelkopf_spectrometer}, and it has since been realized experimentally.\cite{Astafiev} 
Here we propose a quantitative method for extracting the moments of the noise contributing to pure dephasing, which can be different than the noise leading to the energy relaxation. In particular, for $t$$<$$1/\omega_{c}$ we have $\chi^{\text{UDD}}(t) \! \sim \! t^{2n+2}M_{2n}$ (see Eq.~(\ref{eq:FUDD_loww})), where $M_{k} \!\! = \!\! \int \omega^{k}S(\omega) d\omega$ is the $k$-th moment of the spectral density. From the moments $M_{k}$ one can, in principle, reconstruct the noise spectrum.
For SE and two-pulse UDD (equivalent to two-pulse CPMG) we get, respectively, $\chi \! \approx \! M_{2} t^{4}/32\pi$ and $\chi \! \approx \! M_{4} t^{6}/1024\pi$. The observation of $\exp(-t^{4})$ and $\exp(-t^{6})$  decays of $W(t)$ for these one and two pulse sequences will be a signature of the presence of finite $\omega_{c}$.

Fulfilling the condition $2n/t > \omega_{c}$ might however be experimentally challenging. As we discussed in Section \ref{sec:realistic}, in reality the pulse time $\tau_{p}$ has a lower bound, and $n$ can also be limited by accumulation of errors in a long and complicated sequence.
Assuming $\tau_{p}$ is the limiting factor, if $2/\tau_{p}\!\!\gg\!\!\omega_{c}$, it is possible for the filter function to ``reach the cutoff'', and then the previous considerations hold. 

\section{Non-Gaussian random telegraph noise.}  \label{sec:RTN}
The comparison between the experiment\cite{Nakamura_PRL02} and theory\cite{Galperin_PRL06} clearly shows that in charge qubits the decoherence can be dominated by coupling to a single classical fluctuator, which is a source of the RTN ($\beta(t) \! = \! v\xi(t)$ with $\xi(t)$ switching between $\pm 1/2$ with rate $\gamma$). 
Two regimes of decoherence can be identified,\cite{Paladino_PRL02,Galperin_PRL06,Bergli_PRB07} the strong (weak) coupling regime in which $g\!\!\equiv\!\!v/\gamma \!\! \gg \!\! 1$ ($g\!\! \ll \!\!1$). For $g \ll 1$ we are in the ``motional narrowing'' regime: the fluctuator is switching so fast that its influence on the qubit is averaging itself out, leading to large $T_{2}$. Furthermore, since on the relevant time-scale the qubit receives a large number of  ``phase kicks'' from the fluctuator (with typical size of $v/\gamma$), the effective noise affecting the pure dephasing dynamics is approximately Gaussian. On the other hand, for $g \gg 1$ one expects short decoherence time with strongly non-Gaussian features in time-dependence of $W(t)$.\cite{Galperin_03,Galperin_PRL06,Falci_PRL05} 

\begin{figure}
\includegraphics[width=8.5cm]{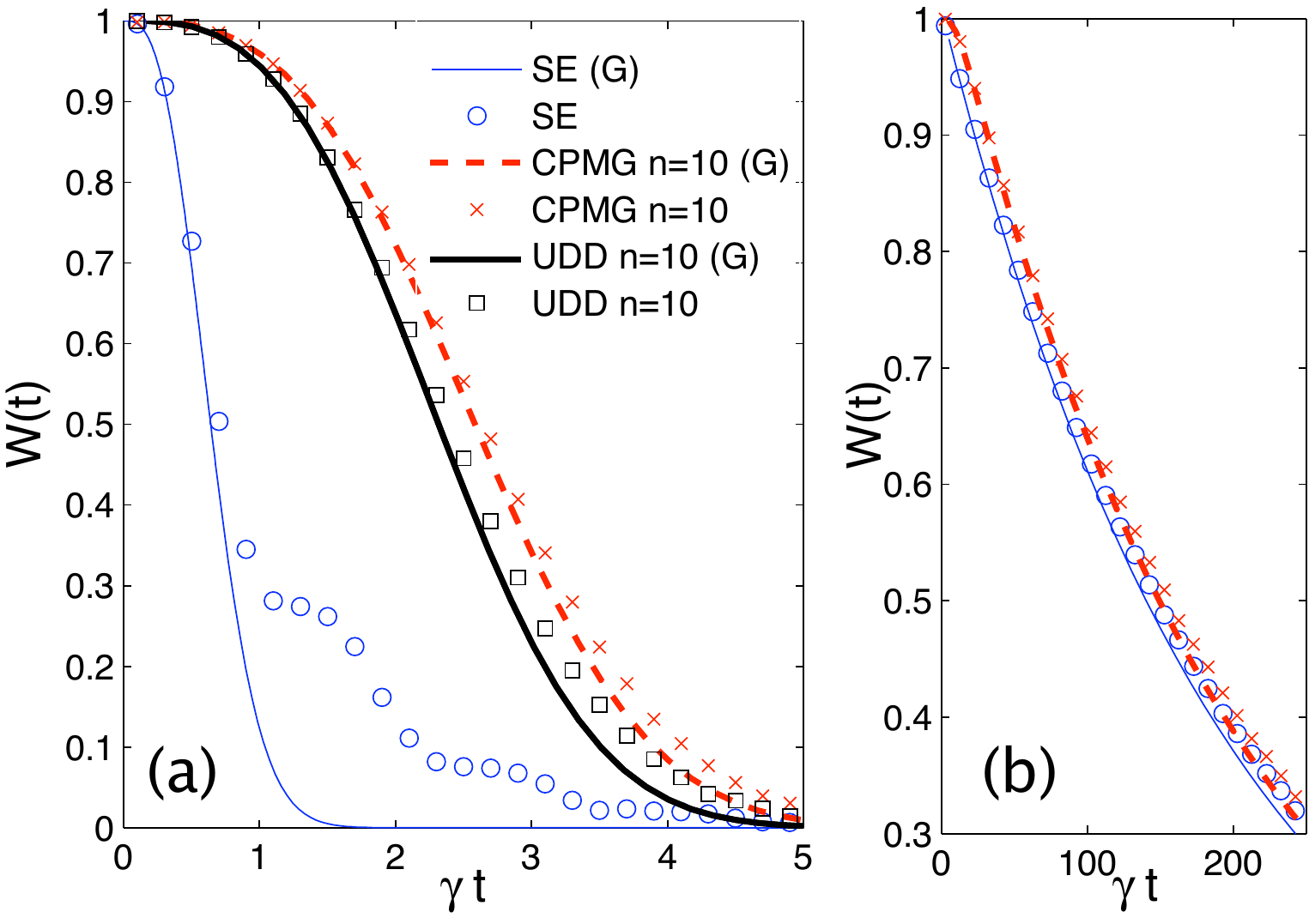}
\caption{(Color online) The decay of qubit coherence $W(t)$ for a single TLF coupled to the qubit for (a) strong coupling $g \! = \! 10$ and (b) weak coupling $g \! = \! 0.2$. The results of the simulation of the RTN are shown as symbols, and the calculations in Gaussian (G) approximation are shown as lines.
CDD$_{4}$ (with $n\! = \! 10$) gives practically the same result as UDD, and thus it is not shown. For  $g \! = \! 0.2$ the Gaussian approximation agrees very well with the exact results. For clarity, only SE and CPMG with $n\!\! = \!\! 10$ are shown in (b). The FID signal for strong coupling (not shown) is an oscillating function for which the SE signal is an envelope.
} \label{fig:RTN}
\end{figure}

We have studied the effect of the pulse sequences on qubit decoherence using  both numerical simulations of the RTN and the Gaussian approximation, in which we plug the Lorentizan first spectral density of the RTN, Eq.~(\ref{eq:Lorentzian}), into Eq.~(\ref{eq:chi_gaussian}). The results for $W(t)$ in both coupling regimes are shown in Fig.~\ref{fig:RTN}. For $g \!\! \ll \!\! 1$, the effect of pulses is marginal, {\it i.e.}~one has to apply a large number of pulses to obtain a visible effect. On the other hand, in the strong coupling regime, application of even a few pulses substantially increases the coherence time. Similar to the case of the Gaussian $1/f^{\alpha}$ noise without a cutoff, the CPMG sequence is the better practical approach.

For $g \!\! \gg \!\! 1$ there are strong deviations from Gaussian behavior in the SE signal (see Fig.~\ref{fig:RTN}a), and the shape of $W(t)$ containing the characteristic plateaus has been derived using various analytical methods.\cite{Galperin_03,Galperin_PRL06,deSousa_PRB03} The values of $v$ and $\gamma$ can be inferred from the position and height of the first plateau.\cite{Galperin_PRL06}
However, one can see in Fig.~\ref{fig:RTN}a that  as we apply more pulses the deviation between the simulation of the exact RTN and the Gaussian approximation decreases. Therefore, with increasing $n$ the simple analytical results following from Eq.~(\ref{eq:chi_gaussian}) become more accurate, {\it i.e.} the non-Gaussian effects are suppressed by pulses.

The explanation of the improvement of Gaussian approximation with increasing $n$ in the strong coupling regime is the following:  the deviation between the exact result for RTN and Gaussian approximation arises from higher order noise correlators in the cumulant expansion of $W(t)$:
\begin{equation}
\ln W(t)  = - \chi^{(2)}(t) - \chi^{(4)}(t) +\,  ...
\end{equation} 
with $\chi^{(n)}\!\sim\! g^{n}$, and $\chi^{(2)}$ given by the expression following from the Gaussian approximation, Eq.~(\ref{eq:chi_gaussian}). The ratio $R(t) \!\! \equiv \chi^{(4)}(t)/\chi^{(2)}(t)$ can be used as a measure of the importance of non-Gaussian effects. We have calculated it for various pulse sequences, finding that $\chi^{(4)}$ is more strongly suppressed by pulses than $\chi^{(2)}$, so that while the coherence time $T_{2}$ is extended with increasing $n$, the time-scale on which the non-Gaussian effects are negligible grows even faster. The details of the calculations are given in Appendix \ref{app:RTN}.
In Fig.~\ref{fig:R} we show that $R(t)$ remains small for a longer time with the application of more pulses. The CPMG sequence is better than UDD at suppressing $R(t)$, which should not be surprising in the light of the fact that UDD is optimized to make only $\chi^{(2)}$ as small as possible. Evidently it suppresses $\chi^{(4)}$ less efficiently than the CPMG sequence.

\begin{figure}
\includegraphics[width=8.5cm]{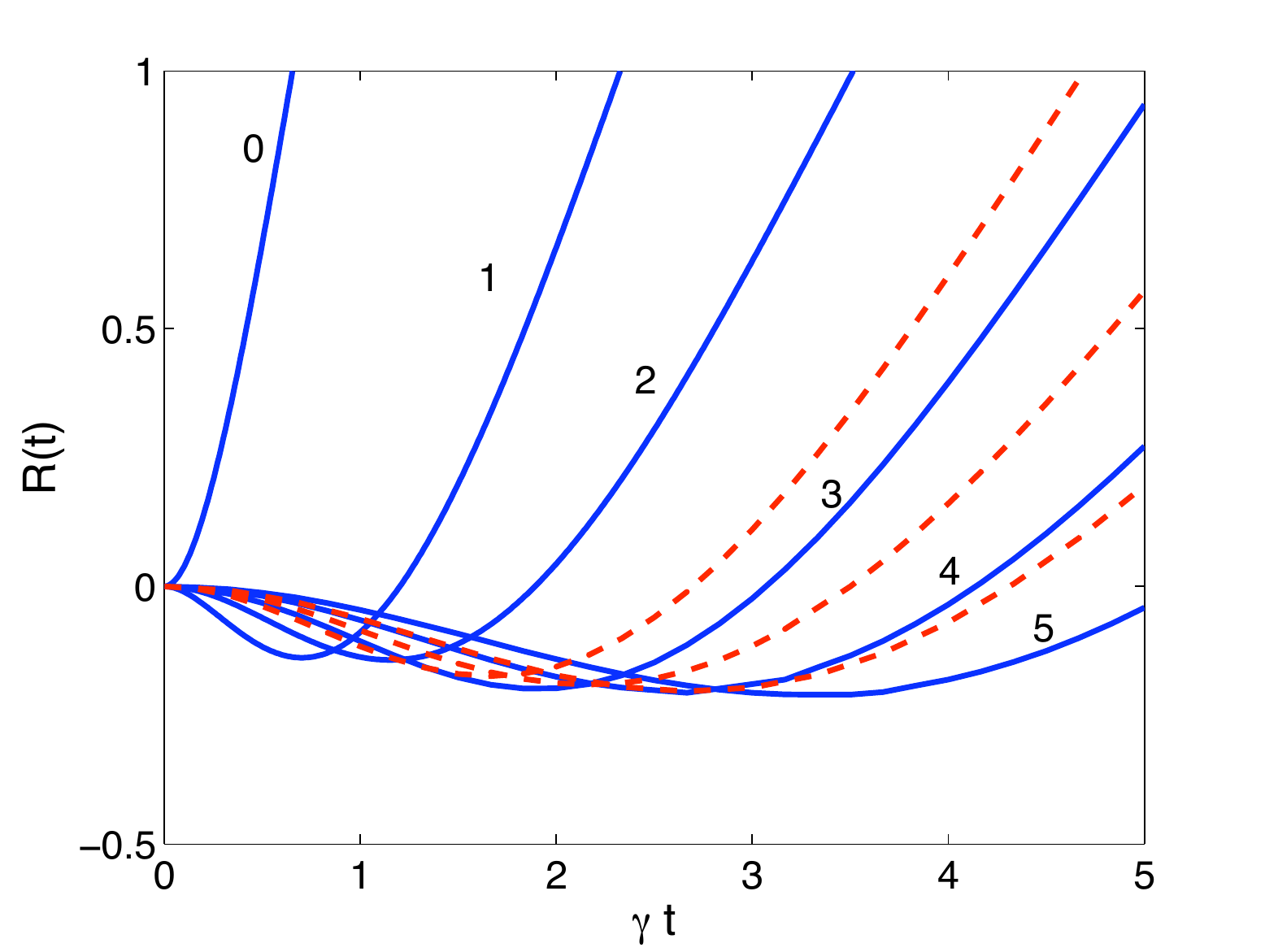}
\caption{(Color online) The ratio $R(t)\! = \! \chi^{(4)}/\chi^{(2)}$ plotted for sequences having $n\! = \! 0...5$ pulses, with RTN characterized by $\gamma\! = \! 1$ and energy of interaction with the qubit $v\! = \! 10$ (so that $g\! = \! 10$).  
The solid lines correspond (from left to right) to FID, SE, and CPMG with $n \! = \! 3,4,5$. The dashes lines correspond to the UDD sequence with $n=3,4,5$. 
With increasing number of pulses, the time at which the non-Gaussian effects start to be quantitatively important  (when $R(t)\! > \! 1$) becomes larger, eventually surpassing the $T_{2}$ time of the qubit.
} \label{fig:R}
\end{figure}

In the Gaussian approach, using the analytical approximations outlined in Section \ref{sec:1f_gaussian} we find that for CPMG in the large $n$ regime we have
\begin{eqnarray}\label{eq:chiRTN}
\chi(t)\! \approx   \!\left\{
\begin{array}{rcl}
&\!& \frac{g^{2}}{24} \frac{(\gamma t)^{3}}{n^2},
\,\,\,\,\,\,\,\,\,\,\,\,\,\,\,\,\,\,\,\,\,\, \gamma t \ll n \,\, , \\\\
&\!& \,\, \frac{g^{2}}{8} ( \gamma t - n),
\,\,\,\,\,\,\,\,\,\, \gamma t \gg n \,\, ,
\end{array}
\right.
\end{eqnarray}
and for UDD  the numerical coefficient in the first equation is larger by a factor of about $1.5$ and $n$ should be replaced by $n+1$ in the denominator.
The first formula holds when the filter function $F(\omega t)$ suppresses the low-frequency ($\omega\! < \! 2\gamma$) flat part of the Lorentzian spectral density, and only the $\gamma v^2/\omega^2$ tail contributes to $\chi(t)$. 
If $g \! \gg \! 1$, most of the decoherence occurs for $\chi(t) \! \sim \!  t^{3}$, and $T_{2}$ falls within this regime. Then the application of more pulses is effective as it decreases the coefficient of $t^{3}$, resulting in
\begin{equation}
T_{2} \approx \frac{2}{\gamma} \frac{n^{2/3}}{g^{2/3}}\,\, , \,\,\,\,\,\,  g\gg 1 \,\,  .  
\end{equation}
On the other hand, for $g \ll 1$, most of decoherence occurs in the long time ($\gamma t\gg n$), $\chi(t) \! \sim \! t$ regime, in which the largest contribution to $\chi$ comes from the flat part of the spectral density. Then, adding a few pulses only prolongs the initial short-time $\chi(t) \! \sim \! t^{3}$ behavior, with marginal effect on the decoherence time ($T_{2}\! \approx \! 8/\gamma g^{2}$). 
However, the initial decoherence ($\gamma t \ll n$) is suppressed as before, resulting in improvement of fidelity at short times.
The pulses affect the $T_{2}$ time only when we apply $n \! > \! 8/g^{2}$ pulses, extending the $t^3$ regime so that $T_{2}$ falls within it. 

\section{Conclusions}
We have analyzed the influence on various pulse sequences on pure dephasing of a qubit affected by classical noise, with emphasis on types of noise relevant for superconducting qubits.
We have shown that successive higher order pulse sequences lead to an improvement of coherence time for both Gaussian $1/f^{\alpha}$ noise and RTN.
We have found that in the presence of a hard ultra-violet cutoff in the Gaussian noise spectrum, the UDD sequence is optimal for suppressing initial decoherence.
However, if one can not ``reach the cutoff'', the CPMG sequence is the best practical approach. This is also true in the case of a single fluctuator coupled to the qubit. There, the application of large $n$-sequences decreases the deviation between exact (non-Gaussian) theory and Gaussian approximation. 
For both $1/f$ noise and RTN we predict substantial practical enhancement in SC qubit coherence under the CPMG pulse sequence. Furthermore, a detailed experimental investigation of the noise mechanisms operational in different SC circuits and samples becomes possible using the UDD pulse sequence, which allows one to gather quantitative information about low frequency noise contributing to dephasing.

\begin{acknowledgments} 
We thank C.J.~Lobb, R.~Schoelkopf, R.W.~Simmonds, L.~Viola, F.C.~Wellstood, and W.M.~Witzel for discussions. This work was supported by the LPS-NSA-CMTC grant and by a fellowship from the Joint Quantum Institute (RL).
\end{acknowledgments}

\appendix

\section{Calculation of the fourth cumulant of the Random Telegraph Noise} \label{app:RTN}
We define the phase $\Phi(t)$:
\begin{equation}
\Phi(t) = v\int_{0}^{t}\xi(t')f(t;t') dt'  \,\, ,
\end{equation}
so that  the decoherence funtion $W(t)$ is given by
\begin{equation}
W(t) = \langle e^{-i\Phi(t)} \rangle \equiv \exp \Big( \sum_{k=1}^{\infty} \frac{(-i)^{n}}{n!}C_{k} \Big)  \,\, ,
\end{equation}
where we have written it using the cumulant expansion.\cite{Kubo_JPSJ62}
The cumulants $C_{k}$ vanish for $k\! > \! 2$ if the statistics of $\Phi(t)$ is Gaussian.
They can be written in terms of moments $M^{\Phi}_{k}(t) \! = \! \langle \Phi^{k}(t) \rangle$, with the first two non-vanishing ones (we assume $\langle \xi \rangle \! = \! 0$, so that the odd-$k$ moments and cumulants vanish) given by
\begin{eqnarray}
C_{2}(t) &= & M^{\Phi}_{2}(t) \,\, , \\
C_{4}(t) &= & M^{\Phi}_{4}(t) - 3[M^{\Phi}_{2}(t)]^{2} \,\, .  \label{eq:C4}
\end{eqnarray}
We also define 
\begin{equation}
-\ln W(t) \equiv \chi(t) = \sum_{k=1}^{\infty} \chi^{(2k)}(t) \,\, .
\end{equation}
In the Gaussian approximation the only non-zero term in the above expansion is $\chi^{(2)}$$=$$\frac{1}{2}C_{2}$. For the RTN the higher-order terms do not vanish, {\it e.g.} we have $\chi^{(4)}$$=$$-\frac{1}{24}C_{4}$. 

In order to calculate $\chi^{(4)}(t)$ we need to understand the structure of the higher-order correlation functions of $\xi (t)$. 
Following Ref.~\onlinecite{Galperin_03} we write $\beta(t)$ as
\begin{equation}
\beta(t) = v \xi_{0} (-1)^{n(0,t)} \,\, ,
\end{equation}
where $\xi_{0}$$=$$\xi(0)$$=$$\pm 1/2$ is the initial condition, and $n(t_{1},t_{2})$ is the random variable giving us the number of flips between times $t_{1}$ and $t_{2}$. From definition of the RTN process we have
\begin{equation}
\langle (-1)^{n(t_{1},t_{2})} \rangle = e^{-2\gamma |t_{1}-t_{2}|} \,\, .
\end{equation}
The two-point correlation function of the noise can be written for $t_{1} \geq t_{2}$ as
\begin{eqnarray}
\!\!\!\! \langle \beta(t_{1})\beta(t_{2}) \rangle \! & = & \! v^{2} \xi_{0}^2 \langle  (-1)^{n(0,t_{1})} (-1)^{n(0,t_{2})} \rangle \nonumber \\ 
\!\!\!\! \!&= & \! \frac{v^2}{4}  \langle (-1)^{n(0,t_{2})} (-1)^{n(t_{2},t_{1})} (-1)^{n(0,t_{2})} \rangle \nonumber \\
\!\!\!\! \!&= & \! \frac{v^2}{4} \langle (-1)^{n(t_{2},t_{1})} \rangle = \frac{v^2}{4} e^{-2\gamma |t_{1}-t_{2}|}  \,\, , \label{eq:f2}
\end{eqnarray}
with the result being the same for $t_{2} \! \geq \! t_{1}$ (since $n_(t_{2},t_{1}) \! = \! n(t_{1},t_{2})$), so that we recover Eq.~(\ref{eq:S_RTN}). In an analogous way we can calculate the four-point correlation function, but now the ordering of time arguments will matter. Assuming $t_{1}\geq t_{2} \geq t_{3} \geq t_{4} $ we get
\begin{equation}
\langle \beta(t_{1})\beta(t_{2})\beta(t_{3})\beta(t_{4}) \rangle = \frac{v^4}{16} e^{-2\gamma(t_{1}-t_{2})} e^{-2\gamma(t_{3}-t_{4})} \,\, . \label{eq:f4}
\end{equation}
Time ordering is crucial here. For any other ordering we have to permute the times on the right-hand side. However, we deal here with multiple integrals of the form
\begin{eqnarray}
\left \langle \big( \int_{0}^{t}d\tau \xi(\tau) f(\tau) \big)^{k} \right \rangle & = &  \nonumber \\
&& \!\!\!\!\!\!\!\!\!\! \!\!\!\!\!\!\!\!\!\!\!\!\!\!\!\!\!\!\!\! \!\!\!\!\!\!\!\!\!\!\!\!\!\!\!\!\!\!\!\!\!\!\!\! \!\!\!\!\!\!\! \int_{0}^{t} dt_{1} ... \int_{0}^{t} dt_{k}  \langle \xi(t_{1}) ... \xi(t_{k})\rangle f(t_{1})...f(t_{k}) \,\, ,
\end{eqnarray}
where we have used the simplified notation $f(t_{i}) \! \equiv \! f(t;t_{i})$.
The integration region (the $k$-cube) can be divided into $k!$ simplexes, each with a definite ordering relation between all the times. 
The integration variables can be relabeled in each integration region, and we obtain
\begin{widetext}
\begin{equation}
\left\langle \big( \int_{0}^{t}d\tau \xi(\tau) f(\tau) \big)^{k} \right\rangle =
k! \int_{0}^{t} dt_{1} \int_{0}^{t_{1}} dt_{2}  ... \int_{0}^{t_{k-1}} dt_{k}  \langle \xi(t_{1}) ... \xi(t_{k})\rangle f(t_{1})...f(t_{n}) \,\, .
\end{equation}
With this formula we get for the moments:
\begin{eqnarray}
M^{\Phi}_{2} & = & \frac{v^{2}}{2} \int_{0}^{t}dt_{1} \int_{0}^{t_{1}}dt_{2} e^{-2\gamma(t_{1}-t_{2})} f(t_{1})f(t_{2})\,\, , \\
M^{\Phi}_{4} & = & \frac{3v^{4}}{2} \int_{0}^{t} \!\! dt_{1} \int_{0}^{t_{1}} \!\! dt_{2} \int_{0}^{t_{2}} \!\! dt_{3} \int_{0}^{t_{3}} \!\!  dt_{4} e^{-2\gamma(t_{1}-t_{2}+t_{3}-t_{4})}  f(t_{1})   f(t_{2}) f(t_{3}) f(t_{4})  \,\, ,
\end{eqnarray}
where the formula for $M^{\Phi}_{2} \! = \! 2\chi^{(2)}$ is simply a different way of obtaining the Gaussian result from Eq.~(\ref{eq:chi_gaussian}) with Lorentzian spectral density. Using Eq.~(\ref{eq:C4}) we obtain the fourth cumulant $C_{4}(t)$.
For small number of pulses $n$, {\it e.g.} for FID, SE, and $n\! = \! 2$ CPMG/UDD (labeled hereafter as $\text{CP2}$) we get
\begin{eqnarray}
C_{4}(t)^{\text{FID}} &= &-\frac{3g^4}{64} \Big( 4\gamma t + e^{-4\gamma t} + e^{-2\gamma t}(4+8\gamma t) -5 \Big) \,\, , \label{eq:C4_FID} \\ 
C_{4}^{\text{SE}}(t) & =& -\frac{3g^4}{64} \Big( 4\gamma t + e^{-4\gamma t} - 8 e^{-3\gamma t} + e^{-2\gamma t}(12-8\gamma t) + 8 e^{-\gamma t} (1+2\gamma t) -13 \Big) \,\, , \label{eq:C4_SE}  \\
C_{4}^{\text{CP2}}(t) & = & -\frac{3g^4}{64} \Big( 4\gamma t( 1+ 2e^{-2\gamma t} - 6e^{-3\gamma t/2} +4e^{-\gamma t} +2e^{-\gamma t/2} ) + \nonumber\\
& &+ e^{-4\gamma t} -8e^{-7\gamma t/2} +24e^{-3\gamma t} -24e^{-5\gamma t/2} -12e^{-2\gamma t} +24e^{-3\gamma t/2} +8e^{-\gamma t} +8e^{-\gamma t/2} -21 \Big) \,\, .
\end{eqnarray}
\end{widetext}
The analytical expressions for larger $n$ become cumbesome, and we resort to numerical evaluation of $C_{4}$. The results for the ratio of the cumulants
\begin{equation}
R(t) \equiv \frac{\chi^{(4)}(t)}{\chi^{(2)}(t)} = -\frac{1}{12}\frac{C_{4}(t)}{C_{2}(t)} \,\, ,
\end{equation}
up to $n\! =\! 5$ are presented in Fig.~\ref{fig:R}.

The fact that with increasing $n$ the higher order cumulants are suppressed more strongly than the Gaussian $C_{2}(t)$ can be understood in the following way. $\chi^{2k}$ is proportional to $2k-$fold time integral of a noise correlation function multiplied by $2k$ functions $f(t_{i})$, each of them alternating between $\pm 1$. Under the multiple integral, and for large $n$, the sign of the product of $f(t_{i})$ switches multiple times, and  with increasing order $2k$ the whole expression is effectively averaged out by the filter functions. 


\begin{thebibliography}{68}
\bibitem[{foo()}]{footnote_optimal}
\bibinfo{note}{Note however the existence of theoretical proposals for quantum
  computer architectures involving only optimally biased qubits,
  Ref.~\onlinecite{Rigetti_PRL05}, and a recent experimental demonstration of a
  two-qubit gate with both qubits at the optimal bias,
  Ref.~\onlinecite{Niskanen_Science07}.}
\end{thebibliography}

\end{document}